\documentclass[12pt]{iopart}

\usepackage{color}  
\usepackage{graphicx}

\begin{document}

\title[Giorgio Parisi]{Giorgio Parisi's scientific portray:\\
Complex Systems and much more}

\author{Leticia F. Cugliandolo}

\address{Sorbonne Universit\'e, Laboratoire de Physique Th\'eorique et Hautes Energies, CNRS UMR 7589, 
4 Place Jussieu, 75252 Paris Cedex 05, France\\
}
\vspace{10pt}
\begin{indented}
\item[]July 2022
\end{indented}

\begin{abstract}
This article summarises the outstanding scientific career of Giorgio Parisi, who was awarded the 2021 
Nobel Prize in Physics, with special emphasis on his contributions to the description  of the equilibrium properties of 
disordered systems.
\end{abstract}

%
%
%
%
%

\section{Introduction}

Giorgio Parisi received the 2021 Nobel Prize in Physics  “for the discovery of the interplay of disorder and 
fluctuations in physical systems from atomic to planetary scales”. The other half was awarded to  Syukuro Manabe 
and Klaus Hasselmann 
“for the physical modelling of Earth’s climate, quantifying variability and reliably predicting global warming”.
These are the official Nobel citations. 

Grouping Giorgio Parisi with the other two recipients  in the same triad may look awkward at first sight. However, a closer look at 
Parisi's impressive scientific production 
shows that one of his  seminal  contributions, the discovery of the stochastic resonance phenomenon, 
is indeed related to climate change. In the original paper,  co-authored with Roberto Benzi, Alfonso Sutera and Angelo Vulpiani,  the  possible 
relevance of this mechanism in the context of climate modeling was already highlighted~\cite{Benzi}.

The stochastic resonance discovery is only one exceptional  contribution in the very long list 
that Giorgio  Parisi has made. One of the 
keywords in the Nobel Prize citation is {\it disorder}. His studies of 
spin-glasses, the paradigmatic disordered system,  led to the identification and interpretation 
of the {\it Replica Symmetry Breaking Ansatz}, which not only solved the standard mean-field spin-glass model, 
but also acted as a source of inspiration to understand many other physical (and not only) systems.

I would describe Giorgio Parisi as a renaissance  researcher with an accordingly wide palette of  interests. 
He has been terribly creative. In particular, in the period going from the mid 70s to the end of the 80s 
he produced many outstanding results which have been highly influential and even opened
full  new lines of research.
He made remarkable contributions in theoretical physics and mathematics, but also in 
computer design, observational methods and data analysis, as well as in interdisciplinary 
fields such as neural networks, combinatorial optimisation, active 
matter and climate science, to name a few. The latter  areas were familiarly called ``the beyond'' by Roman students, 
in reference to the book {\it Spin-glass theory and beyond}, by Marc M\'ezard, Giorgio Parisi and Miguel \'Angel Virasoro~\cite{MePaVi}, which marked an epoch. 
Many of these problems belong to the nowadays called {\it Complex Systems}, the focus of this journal, 
and I will describe them in some detail in the body of this article.  Clearly enough, it would not be possible to cover 
all the topics he worked on in a single article with restricted length, so I will  make a selection, 
which tells just one of the many possible stories of his scientific career.

I close this introduction with a short description of Giorgio Parisi's academic {\it vitae}.
Giorgio Parisi studied physics at Sapienza University of Rome and obtained his 
Laurea in 1970. For the next 10 years he held a researcher position at the Laboratori 
Nazionali di Frascati (1971 - 1981), and during this period he was
visiting scientist at Columbia University, New York, USA (1973 - 1974), 
the Institut des Hautes \'Etudes Scientifiques \`a Bures-sur-Yvette (1976 - 1977) and \'Ecole Normale Sup\'erieure de Paris (1977 - 1978)
both in France. In 1981 he became full professor at the University of Rome Tor Vergata,  and in 1992 he moved to 
Sapienza University of Rome as full professor in Theoretical Physics. Since 1987 he is a member of the Accademia Nazionale dei Lincei, 
which he presided during the period 2018 - 2021. Presently, he is the vice-president of this Academy, Emeritus professor at Sapienza,
and member of the French and American Academy of Sciences. His list of publications includes one thousand papers, 
approximately. He has written the first book devoted to 
Field Theory and Statistical Physics~\cite{Parisi-Field-Theory}, and 
authored and co-authored several other books.

\section{Roma, the environment}
\label{sec:Roma}

\begin{figure}[h!]
\centerline{
\includegraphics[scale=0.15]{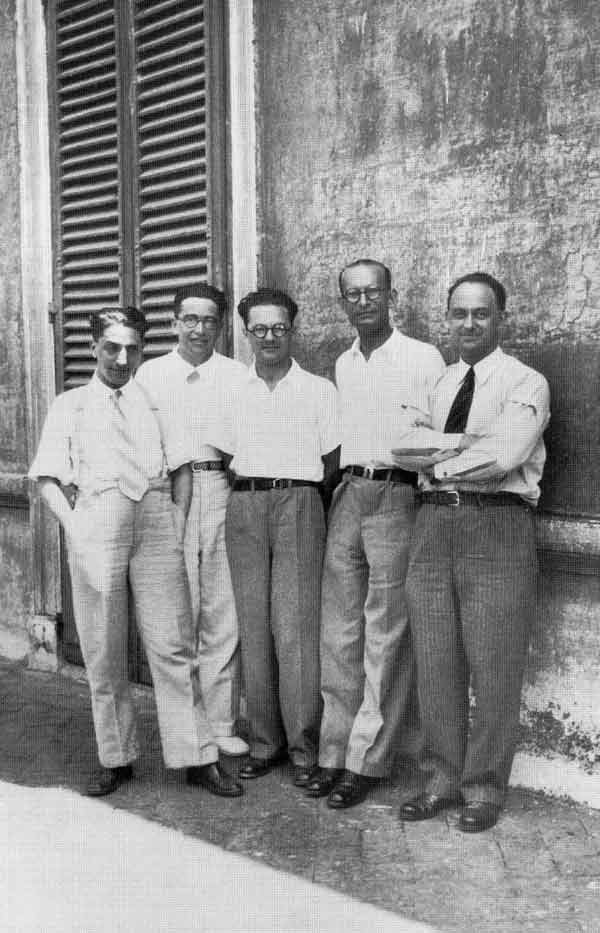}
\includegraphics[scale=0.11]{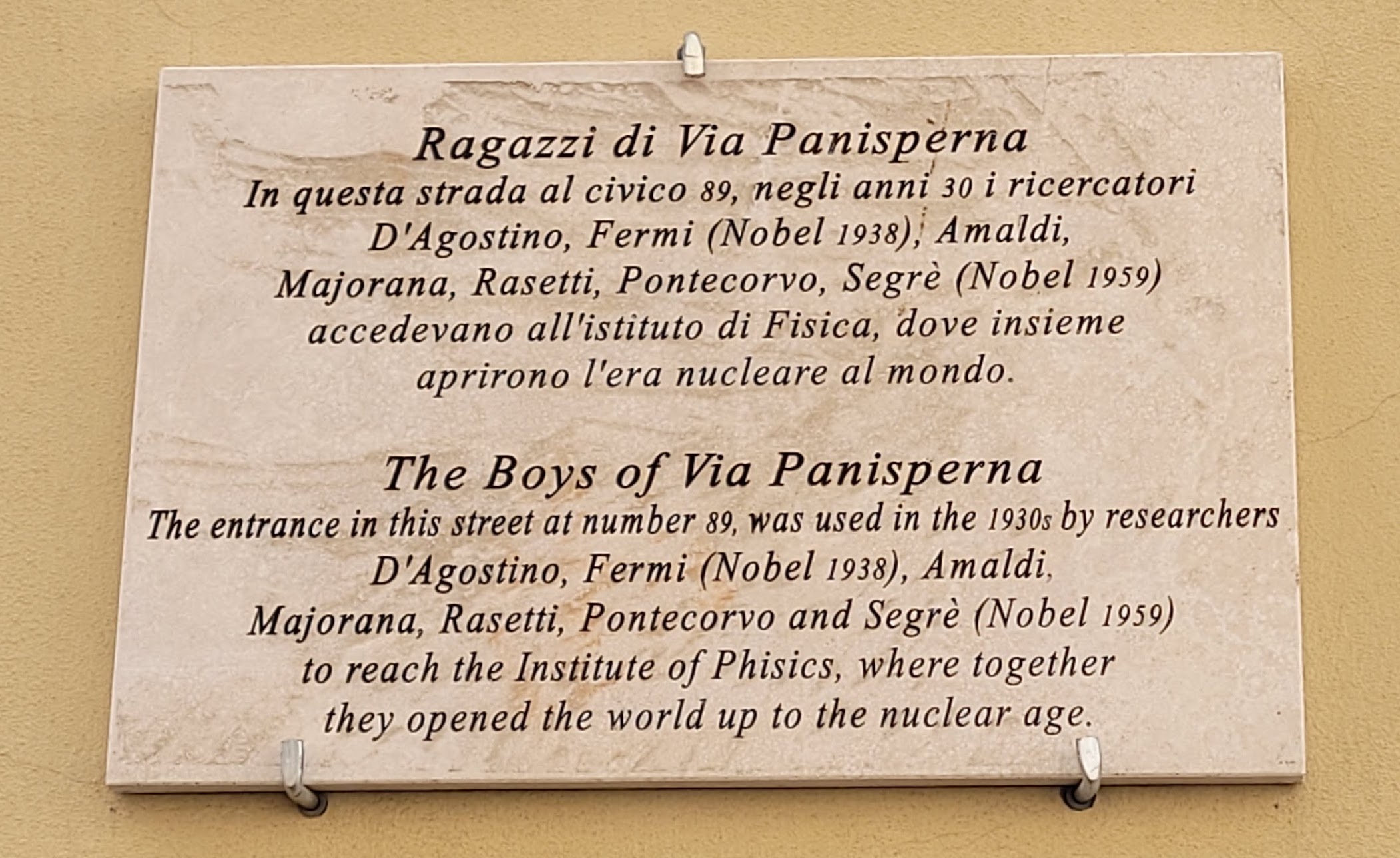}
}
\caption{Picture of the ragazzi Panisperna taken from Wikipedia and my photo of the plate
standing on the wall of the site.}
\label{fig:Fermi}
\end{figure}

Roma has a long tradition of excellence in Physics, and even more so the theoretical aspects of it. 
The modern period started with the ``Ragazzi di via Panisperna" 
group at the Regio Istituto di Fisica dell'Universit\`a di Roma, led by Enrico Fermi, Fig.~\ref{fig:Fermi}. In the early 30s
the group's research interests moved from atomic to nuclear physics~\cite{Segre} with, 
notably, the development of the theory of $\beta$ decay, and the discovery of 
artificial radioactivity by neutron bombardment and slow neutrons. 
Although Fermi had to leave Italy with the advent of fascism, his school managed to survive beyond WWII
mainly due to the efforts of Edoardo Amaldi who was the only ones who stayed in Rome.

Getting closer to Parisi's student period, the Physics Department 
at Sapienza was populated by very influential young professors like Giovanni Jona-Lasinio, 
Carlo di Castro, Giovanni Gallavotti in the statistical physics, condensed matter
and mathematical physics areas, and Nicola Cabibbo and Guido Altarelli in the particle physics one, for example.
These people, and others, formed  an amazing professorial cohort.

\begin{figure}[h!]
\centerline{
\hspace{1cm}
\includegraphics[scale=0.5]{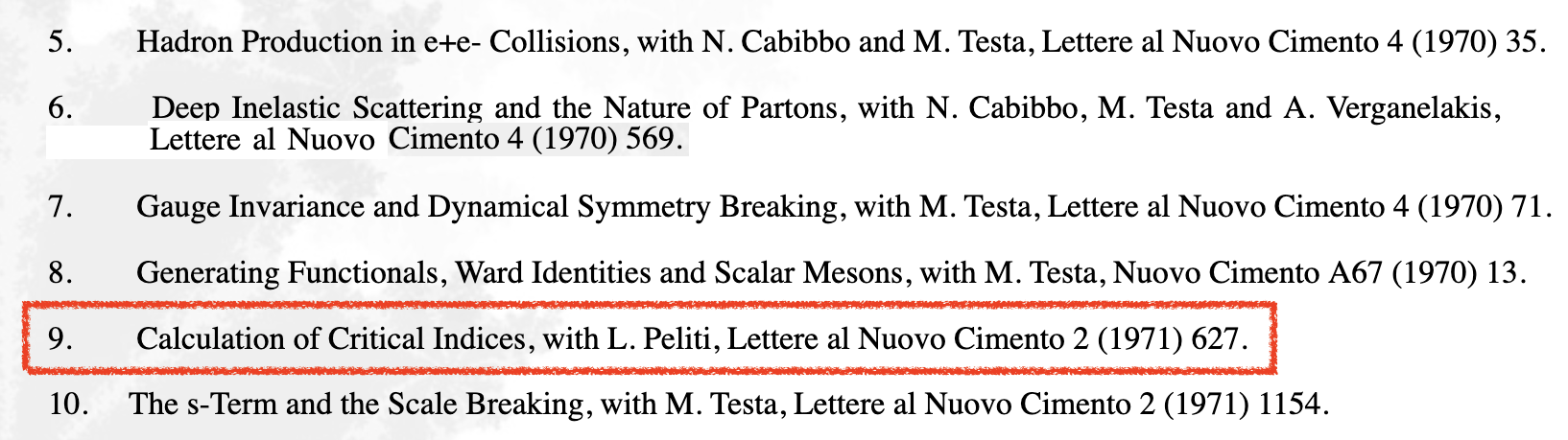}
}
\caption{List of papers. Annotated screenshot from G. Parisi's webpage
{\tt https://chimera.roma1.infn.it/GIORGIO/papers.html}}
\label{fig:early-papers}
\end{figure}

Parisi joined Cabibbo's group and he did his Laurea Thesis under his supervision. His initial works were on 
particle physics. Six of his early papers are listed in Fig.~\ref{fig:early-papers}. However, looking closer at this list, one can spot the 
article {\it Calculation of Critical Indices},  co-authored with his former
classmates, Luca Peliti and Marco D'Eramo, in which closed equations for the critical exponents of the 
$\lambda$-point of Bose liquids were found~\cite{Deramo-Parisi-Peliti, Parisi-Peliti}. 
In a sense, these papers appear as an initial interest in statistical physics problems.
Other works of that time dealt with the conformal group, the two-dimensional conformal anomaly,
and the nowadays reborn conformal  bootstrap 
method to study phase transitions (see, {\it e.g.}~\cite{Ferrara}), which was proposed in the early 70s by Alexander Polyakov
in the USSR, 
but also Raoul Gatto (another prominent figure in Italian theoretical physics who trained numerous ``gattini'') and associates, 
independently~\cite{Polyakov,Gatto}.

In this early period Parisi was trying to find methods and understand field theories with strong
interactions and he was attacking this problem from two ends, the high energy physics  and the 
phase transitions ones (for his personal recollection see~\cite{Parisi} in~\cite{interviews}).

\section{Paris, so many collaborations}

Parisi spent two very fruitful years in France where he continued working on particle physics but not only. 
His article with Guido Altarelli (who was at Ecole Normale Sup\'erieur at the time),
{\it Asymptotic freedom in parton language}, developed the nowadays so-called
Dokshitzer-Gribov-Lipatov-Altarelli-Parisi evolution equations which describe the variation of the 
parton distribution functions with varying energy scales~\cite{Altarelli-Parisi}. 
This work, his most cited paper, was central to  the 2015 EPS High Energy and Particle Physics Prize 
for pioneering research on the structure of protons and
“for having developed the scheme of a probabilistic field theory for the dynamics of quarks and gluons, enabling a 
quantitative understanding of high-energy collisions between hadrons”. 
As Giorgio mentions in his {\it Historical and personal recollections of Guido Altarelli}~\cite{Parisi-Altarelli}, 
``...Guido liked to remark that it is the most cited French paper in the field of high energy physics."

In parallel, Parisi collaborated with \'Edouard Br\'ezin, 
Claude Itzykson and Jean-Bernard Zuber, 
from the Service de Physique Th\'eorique de Saclay,  on the use of matrix models in their infinite size limit, 
an idea pioneered by Gerard 't Hooft, to count planar diagrams (or ``maps'' in mathematical language)~\cite{Brezin}. 
Beyond a physics oriented technique that allows one to recover many results on planar graphs 
previously derived by the celebrated mathematician 
William Tutte with combinatorial methods, this paper introduced a number of tools and 
ideas which became very fruitful later: the large $N$ saddle-point method, the relevance of the density of 
eigenvalues to pin-point the origin of singularities,  {\it etc.}

Many of his long-lasting collaborations involve researchers from French Institutions. 
The seeds  were sown in this period.

\section{Spin glasses}

In the late 70s, calculation tricks in which a discrete parameter was made continuous and its  limit 
towards a convenient (but sometimes strange) value was taken were in the air. 
In 1972 such methods were independently used in the 
statistical physics and particle physics context.
 de Gennes had put forward the $n\to 0$ limit of the $O(n)$ model to obtain the statistics of polymers in dilute solutions~\cite{deGennes}. Bollini and Giambiagi~\cite{Bollini}, and independently 
't Hooft and Veltman~\cite{tHooftVeltman}, had proposed to make the dimension of space a complex parameter and thus 
regularise otherwise diverging integrals appearing in Feynman diagrams of field theories. Yet another example, though not fully resolved at the time, 
was the application of the replica method to study spin-glasses. The resolution of this problem was one of Parisi's main achievements.

\subsection{The material and its models}

The archetypical physical realisation of a {\it disordered} system is a spin-glass. These are magnetic alloys in which magnetic impurities
are placed at fixed random positions with a given concentration, both determined by preparation~\cite{Mydosh}. Pairs of magnetic moments
(henceforth spins) interact {\it via}  RKKY exchanges.  The quenched random positions induce quenched random interactions 
since the exchanges depend on the distance between the spins, in a way in which they can have both 
signs and decay with distance as a short-ranged power law. The presence of ferromagnetic and anti-ferromagnetic interactions
leads to  {\it frustration}, in other terms, the fact that each spin can receive contradictory messages from its neighbours. 
The dimension of the spins depends on the material details; they can have three components 
(Heisenberg), two components (XY) or only one (Ising) with fixed 
modulus in all cases. Experiments showed a kink in the magnetic susceptibility, suggesting a phase transition
towards a low temperature phase of unknown kind.

Edwards and Anderson (EA)~\cite{Edwards-Anderson} simplified the modelling and considered that the 
$i=1, \dots, N$ spins sit at the vertices of a regular three dimensional lattice while the fixed couplings are
drawn from a probability distribution, typically Gaussian with zero mean (if there is no 
ferromagnetic nor anti-ferromagnetic bias) and finite variance. 
EA also proposed a dynamic  order parameter as the long time limit of the time-delayed self-correlation.
In a static calculation one gives the name of EA order parameter to
\begin{equation}
q_{\rm EA} = N^{-1} \sum_{i=1}^N m_i^2 \qquad\qquad \mbox{with} \qquad\qquad m_i = \langle s_i\rangle
\; , 
\end{equation}
which, at zero magnetic field, should vanish above a critical temperature and be different from zero below it, where the spins should 
acquire a local magnetisation varying in space and with both signs.

The peculiarities that a 
single sample may have are washed  away by the {\it self-averaging} of the (disorder dependent)  
free-energy density $f_J$
and the relevant global observables that one can derive from it. Self-averaging states that the typical 
behaviour coincides with the disorder averaged one. This fact is favourable in the sense that 
one does not need to focus on each single sample separately, but brings about the difficulty
of having to calculate the disorder average of the logarithm of the partition function to later derive 
from it the observables. 

\subsection{The replica method}
\label{subsec:replicas}

Following a pioneering paper by Robert Brout~\cite{Brout},
Edwards and Anderson pointed out that the {\it replica trick} could be used to represent the logarithm 
with a Taylor expansion, and transform the calculation in 
\begin{equation}
-\beta [f_J]  = \lim_{N\to\infty} \frac{1}{N} \, [ \ln {\mathcal Z}_J] =  \lim_{N\to\infty} \frac{1}{N}  \lim_{n\to 0} \frac{[{\mathcal Z}^n] -1}{n}
\; , 
\label{eq:replica}
\end{equation}
where the square brackets represent the average over the couplings $J_{ij}$ weighted with their distribution. In practice,
it is still not possible to do this calculation for the EA model.

The mean-field version of the EA model is due to Sherrington and Kirkpatrick (SK)~\cite{Sherrington-Kirkpatrick}. It
consists in placing the spins on a fully connected graph with no notion of 
distance, and scaling the variance of the couplings $J_{ij}$ with the number of spins 
so as to ensure an interesting thermodynamic limit. The replica method (\ref{eq:replica}) 
can now be applied until a later stage in which an $n\times n$ matrix $Q$
with elements $Q_{ab}$ and $a,b=1, \dots, n$ appears as an auxiliary element in the calculation.
In the thermodynamic limit $N\to\infty$ taken {\it before} the $n\to 0$ limit (that is exchanged with it), 
the elements of this matrix become the expected values of the {\it overlaps} between the 
replicas 
\begin{equation}
Q_{ab} = [\langle \frac{1}{N} \sum_{i=1}^N s_i^a s_i^b \rangle]
\; , 
\end{equation} 
at the saddle-point level. Quite naturally, SK 
assumed that the replicas where indistinguishable and just a calculation convenience, and 
set the non-diagonal elements to be all equal, $Q_{a\neq b}=q$, coining the so-called {\it replica 
symmetric Ansatz}. The calculation simplified considerably and the saddle-point equation for $q$ 
was just a slight modification of the well-known Curie-Weiss one for the magnetisation in a 
pamagnetic-ferromagnetic transition. Unfortunately,  the final solution 
exhibited two inconsistencies: the zero temperature entropy was negative~\cite{Sherrington-Kirkpatrick} 
and the saddle-point was not stable~\cite{deAlmeida-Thouless}. 

Following SK, several researchers tried to break the replica symmetry and
hopefully thus solve the two problems mentioned above.
Notable proposals came from Andr\'e Blandin (Laboratoire de Physique des Solides 
d'Orsay)~\cite{Blandin} and Alan Bray and Mike Moore (Manchester)~\cite{BrayMoore}
but were not fully correct. (See Marc Gabay's and Mike Moore's recollections of 
this hectic period~\cite{interviews}.)  

\begin{figure}[h!]
\centerline{
\hspace{2cm}
\includegraphics[scale=0.5]{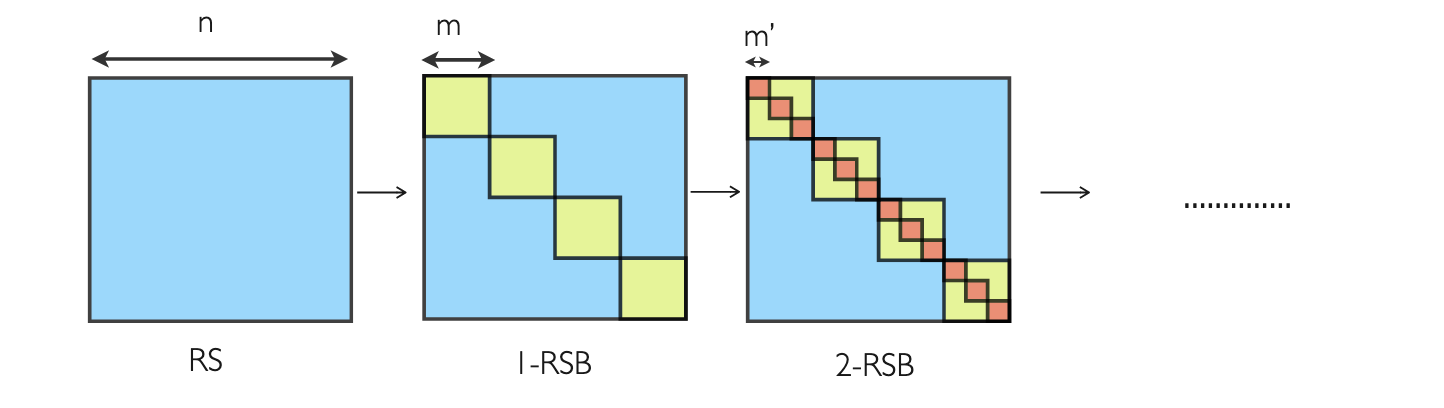}
}
\caption{Sketch of the (discrete) replica symmetry breaking scheme. Figure adapted from~\cite{Morone}.}
\label{fig:replica}
\end{figure}

Parisi came to know  this problem, 
considered it an interesting mathematical challenge (in his own words), 
and found the  {\it replica symmetry breaking Ansatz} with no un-physical issues~\cite{Parisi1,Parisi2,Parisi3}.
His scheme is hierarchical, with the matrix $Q$ being made of diagonal boxes within diagonal boxes each with 
different sizes and elements $q_0<q_1< \dots < q_{\rm EA}$,
see Fig.~\ref{fig:replica}, all determined by the saddle-point method (the various sizes $m$ are not the magnetization and do not 
take integer values). For the SK model this process is repeated with no end and it is 
taken to the continuum limit. This is the {\it full} replica symmetry breaking scheme. In $p>2$ spin models the process ends at the first step, 
and a {\it one step} replica symmetry breaking is enough.

The outcome of the rather cumbersome, but well posed, calculation of $[f_J]$ is basically reduced to the 
determination of the {\it overlap distribution function}, which is the order parameter of these problems, 
\begin{equation}
P(q) =  \delta(q-Q_{ab}) 
\; . 
\end{equation}
In the high-temperature phase this is just a delta function at zero.
In the low-temperature phase, $P(q)$ turns out to have 
two delta peaks at $q=\pm q_{\rm EA}$
but also weight at smaller absolute values.
The precise form of $P(q)$
depends on the model. One basically finds three classes:  
\begin{itemize}
\item[-] Ferromagnetic with no further peaks, $q_{\rm EA} = m^2$ and $m$ the magnetisation density. 
\item[-] Structural glass like with just another delta peak at $q=0$.
\item[-] Spin-glass like with non-zero weight for all  $|q|<q_{\rm EA}$.
\end{itemize}
The first two classes are realised by the SK model  with spherically constrained spins, instead of Ising, and the 
extension to multi (more than two) body random interactions of the Ising and spherical models, respectively. The latter is the 
SK model one.

The picture that emerged from the replica calculation is that the SK model has an infinite number of equilibrium states 
in the thermodynamic limit. In terms of the local magnetisations, each of these states is characterised by its own
ensemble of $\{m^\alpha_i\}$, with $\alpha$ labelling the state. The overlap between two states 
is $q_{\alpha\beta} = N^{-1} \sum_{i=1}^N m_i^\alpha m_i^\beta$ and the EA parameter 
$q_{\rm EA} = q_{\alpha\alpha} $ for any $\alpha$~\cite{DeDominicis}. 
(More about the probabilities $p_\alpha$ will be explained in Sec.~\ref{subsec:TaP}.)
The disorder dependent order parameter should then be
$P_J(q) = \sum_{\alpha\beta} p_\alpha p_\beta \, \delta(q-q_{\alpha\beta})$, with $p_\alpha$ the 
probability of  state $\alpha$,  and its average $P(q) = [P_J(q)]$, recovering in this way the connection 
with the replica order parameter.
 
 The next step towards the full understanding of the SK model was the 
identification of the {\it ultrametric} organisation of its equilibrium states~\cite{Toulouse}.
This peculiar structure means that for any three states chosen at random,  two overlaps are equal and smaller than the third one.
\begin{equation}
q_{\alpha\beta} = \min(q_{\alpha\gamma}, q_{\beta\gamma}) \;\; \mbox{or} \;\;
q_{\alpha\gamma} = \min(q_{\alpha\beta}, q_{\beta\gamma}) \;\; \mbox{or} \;\;
q_{\beta\gamma} = \min(q_{\alpha\beta}, q_{\alpha\gamma})
\; . 
\end{equation}
A graphical representation of this structure gives rise to what is called {\it Parisi's tree}. In the spherical 
$p>2$ model the states are orthogonal in the sense that their overlap vanishes and this gives rise to the 
delta peak at $q=0$ in $P(q)$.

\subsection{Numerical tests}

In numerical simulations one does not have access to the replicas of the analytic calculation but to real replicas 
created as copies of the system with the same coupling strengths but independently evolving spins ({\it e.g.} with the Monte Carlo
rule). The overlaps are then computed as the correlation between the spin configurations of the different copies
being careful about reaching thermal equilibrium. 
Such  simulations of the SK model exhibit
a multi-peak structure of $P_J(q)$ which demonstrates the existence of several equilibrium states with different properties~\cite{Marinari}. Those  
overlaps are not saturated in the sense that $q$ is smaller than $q_{\rm EA}$ which is interpreted as the 
overlap of two configurations in the same equilibrium state. Moreover, the average over many realisations of 
disorder of $P_J(q)$ suggests that the $[P_J(q)]$ does indeed behave as $P(q)$ in Parisi's solution.

The RSB solution strictly applies to the equilibrium properties of mean-field disordered models. Whether it
also applies to the finite dimensional Edwards-Anderson model has been a topic of debate for many 
years. Without entering into the technicalities of it, I want to mention that the Roman Statistical Physics 
group and collaborators, namely in other Italian physics departments and in Spain, made an enormous effort
to put these ideas to the numerical test. Special purpose computers were built to this effect and I will 
describe these projects in Sec.~\ref{sec:computers}. 

\subsection{Mathematical proofs}

For long, the various theoretical physics passages performed in Parisi's solution of the SK model were not fully accepted by 
part of the community.  Strikingly, Francesco Guerra, from Sapienza University, showed that Parisi's formula for the SK $[f_J]$ 
is a lower bound of the exact one~\cite{Guerra}, while  a bit later Michel Talagrand, from Sorbonne
Universit\'e, proved that it is an upper bound~\cite{Talagrand}. Therefore, Parisi's expression is the exact one. 
These proofs use completely different methods and do not 
go through the use of replicas. The agreement between their $[f_J]$ and the one derived from the replica method lifted 
all doubts about the correctness of the latter, when applied to mean-field models. 
Furthermore, Dmitry Panchenko also found a way to justify the ultrametric property with 
mathematically justified techniques~\cite{Panchenko}. 

\subsection{Complex landscapes \& dynamics}
\label{subsec:TaP}

Disordered systems are usually associated to complex landscapes. The latter are the extension of the Ginzburg-Landau 
order-parameter dependent free-energy function(al) to cases in which there is an extensive number of ordered parameters 
to care about.  The free-energy function(al) with equilibrium (and metastable) states as global minima (or local saddle-points)
is the one introduced by Thouless-Anderson-Palmer (TAP)~\cite{TAP}, $f_{\rm TAP}(\{m_i\})$, and it determines the landscape. 
The classification in three classes deduced from the replica calculation is 
confirmed by the analysis of these landscapes in the way explained by Cirano De Dominicis and A. Peter Young~\cite{DeDominicis}.
Basically, the state probabilities $p_\alpha$ (already mentioned in Sec.~\ref{subsec:replicas}) are associated to Boltzmann weights, 
$e^{-\beta f_\alpha}/Z$, with free-energy $f_\alpha = f_{\rm TAP}(\{m_i^\alpha\})$.
The calculation of $P_J(q)$, or of other quantities like the physical disorder-averaged free-energy, 
involve as a much relevant ingredient: the {\it complexity} or {\it configurational entropy}, that is, the logarithm of the number of 
extrema at each free-energy (continuous) level $f$.

The geometric properties of these landscapes have important consequences on the thermodynamic and dynamic properties. 
Spherical $p>2$ spin models~\cite{Crisanti} present a marginally stable threshold level, lying 
higher than (exponentially many) stable metastable and equilibrium states. Below a critical temperature, the threshold 
acts as an attractor of the dynamics 
following a high temperature quench~\cite{CuKu93}. These models have a very similar phenomenology 
to the one of fragile glasses and provide a mean-field description of them. In contrast, in the SK model 
no such structure stops the  relaxation.
The asymptotic evolution has many points in common with the equilibrium properties like, for example, 
the replacement of the ultrametric arrangement of equilibrium states by an ultrametric organisation of two-time 
dependent correlation functions~\cite{CuKu94};
\begin{equation}
\lim\limits_{
t_3 \to \infty}
C(t_1,t_3)
= 
\min(C(t_1,t_2), C(t_2,t_3))
\qquad\qquad
t_1 \gg t_2 \gg t_3
\; . 
\end{equation} 
One has to notice, however, that the out of equilibrium 
relaxation occurs in a different region of the configurational space.

The equilibrium $P(q)$ has a dynamic counterpart~\cite{CuKu93}
\begin{equation}
P_d(C) = \int_0^C dC' \, X(C') \qquad \mbox{with} \qquad X(C) = \lim\limits_{\stackrel{t,t' \to \infty}{C(t.,t') = C}} \; \frac{T R(t,t')}{\partial_{t'} C(t,t')}
\; . 
\end{equation}
Here, $C$ is the two-time self-correlation and $R$ the two-time linear response measured at time 
$t$ to an infinitesimal perturbation applied at time $t'$. None of them is stationary in the low temperature phase of
the infinite size model, since the equilibration time diverges in the thermodynamic limit and it is unable to reach equilibrium.
An $X(C)$ different from one demonstrates a violation of the equilibrium fluctuation-dissipation theorem.
It can be interpreted as as the ration between the bath temperature and a 
(scale dependent) effective temperature $X(C) T/T_{\rm eff}(C)$~\cite{Cukupe}.
In the SK model the functional form of $P_d(C)$ is identical to the equilibrium one $P(q)$. In spherical $p>2$ models, 
the global structure (with two peaks apart from the spin reversed one) is the same but the value of the dynamic 
$q_d = \lim_{t-t'\to\infty} \lim_{t'\to\infty} C(t,t')$ (the actual order parameter defined by Edwards \& Anderson) 
is different from the static one derived with the replica method. This feature is one of the indications of the 
fact that the non-equilibrium relaxation takes places in a region of phase space (the threshold) which is 
very different from the one of equilibrium states. It is also one of the reasons why systems with $p>2$ are
considered to be the mean-field models of fragile glasses.
The formal relation between the static replica overlaps and order parameter, and the dynamic correlations and 
fluctuation-dissipation violation, observed in the SK model, was claimed to carry over to the finite dimensional 
SK model in~\cite{FMPP} using a stochastic stability argument.
 
\subsection{The Beyond}

The power of the replica trick and the replica symmetry breaking {\it Ansatz} were soon recognised by researchers 
working in interdisciplinary fields, notably problems belonging to the areas of biophysics and computer science. For instance, the 
evaluation of the maximal capacity of Hopfield neural networks, that is which is the maximal number of patterns that the network can store
and retrieve, and how it depends on the number of neurons, was almost immediately performed with this technique~\cite{Amit}. 
Similarly, the use of the replica method to identify phase transitions in random optimisation problems were 
also quick~\cite{Vannimenus,MePa,FuAnderson}. In this context, the existence of phase transitions in computational problems 
were shown with the statistical physics methods of disordered systems~\cite{Monasson}. Many other 
applications can be found in the ``Beyond'' book~\cite{MePaVi}.

Of particular importance was the more recent analysis of the paradigmatic Combinatorial Optimisation problem,  
K-satisfiability, with ideas stemming from spin-glass theory. 
Such  problems involve $N$ variables which must satisfy $M$ constraints. 
In the case of the K-sat problem, each of these constraints is  the validation of (at least) one requirement on the values taken by 
the K Boolean variables involved. The computer science tasks are, for example,  to find the $N$ variables' assignment(s)  which 
satisfy all constraints,  to compute the number of ways in which these constraints can be verified, and from it
the complexity, and to devise algorithms to most efficiently - that is, with the fewest operations - find these assignments.

In their hardest realisations K-sat (with K $>2$) is supposed to be NP complete. In its random version, ensembles of problems are studied 
statistically. A concrete algorithm built as an extension of the cavity method ({\it belief propagation}),  which now takes into account the 
existence of many states ({\it survey propagation}), was proposed by M\'ezard, Parisi and Zecchina to attack randomly generated instances
of K-sat for parameters such that they are hard to solve~\cite{MePaZe}. These authors received the Onsager prize of the American 
Physics Society in 2016 for their achievement.

Even more recently, the  similarity between resource-competition models and continuous constraint satisfaction problems in their convex regime
was exploited to study the former with the replica method. The transition between a ``shielded" phase, where a collective and self-sustained behavior emerges, and a ``vulnerable" phase, where a small perturbation can destabilize the system and contribute to population extinction was 
exhibited~\cite{Altieri}. Other ecosystems models, like the random Lotka-Volterra one, are currently being studied with this technique.

\subsection{Applications to structural glasses}

The replica theory methods have been extended and applied to particle systems in recent years by Parisi and collaborators.
The {\it effective potential} technique, devised with Silvio Franz, has proven to be a very useful tool to find the 
dynamic critical temperature with purely static methods adapted to deal with this question~\cite{FranzParisi}. 
More recently, 
the replica method was very successfully applied to the study of systems of particles in interaction in infinite dimensional~\cite{Urbani}.
Important issues like the limits of validity of the mode-coupling approximation could thus be quantitatively addressed.

\section{Stochastic processes}
\label{sec:stochastic-processes}

\subsection{Stochastic quantisation}

The stochastic quantisation technique~\cite{Parisi-Wu} proposes to make 
the fields of interest depend on an additional fictitious real time and
evolve them with a  Langevin equation  tailored so that, in the asymptotic limit $t\to\infty$ limit, the field 
probability distribution is the one of the equilibrium quantum field theory of interest (with the Euclidean action). 
In a sense, stochastic quantisation is the continuous 
time parallel of the quantum Monte Carlo technique. Soon after the introduction of this method by Parisi and Wu, 
many quantum field theories were studied in this way and a review article collecting some of these applications 
is~\cite{Damgaard-Huffel}. 

\subsection{The Kardar-Parisi-Zhang equation}


The Kardar-Parisi-Zhang (KPZ) equation for kinetic surface growth~\cite{Kardar} (directed, with no overhangs) 
generalises the Edwards-Wilkinson (EW) linear one.  In the latter, the local velocity of the interface height $h$ with respect to a 
$d$-dimensiaonl substrate 
with positions $\vec x$  is determined by the 
local elastic force and white thermal noise only. KPZ added a non-linear force proportional to the square of the local
gradient of the surface:
\begin{equation}
\underbrace{
\partial_t h(\vec x,t) 
\ \
=
\ \
\nu \nabla^2 h(\vec x,t) 
}_{\rm Edwards-Wilkinson \; (EW)}
\textcolor{black}{ 
\ + \
\underbrace{ 
\lambda (\vec \nabla h(\vec x,t))^2
\ 
}_{\rm KPZ}
}
\
+ 
\underbrace{\; \xi(\vec x,t) \; }_{\rm EW \; white \; noise}
\hspace{0.75cm}
\raisebox{-0.5cm}{
\includegraphics[scale=0.125]{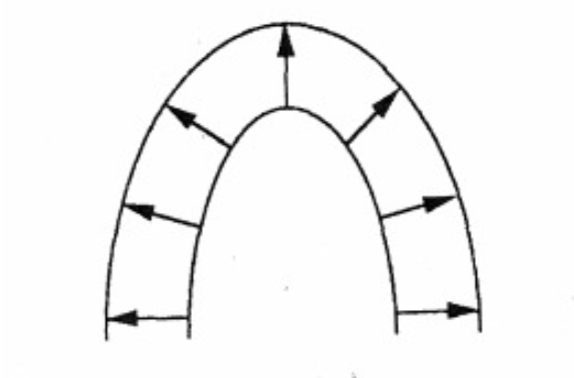}
}
\end{equation}
see the sketch on the right of the equation.
The added term changes dramatically the behaviour of $h$ which is no longer a Gaussian field and presents
a highly non-trivial kinetic roughening. The 
universality class of the growth is determined by, for example, the exponents characterising the scaling properties of the 
surface roughness, and is different from the EW one. This new universality class describes a broad range of non-equilibrium fluctuations, 
beyond those of growing interfaces. Some cases in which dynamic fluctuations have been found to belong
to the KPZ universality class are directed polymers and particle transport, the interfaces of bacteria colonies on agar, 
the slow combustion of paper, and the growth of solid thin films. New examples continue to appear.

It is relatively easy to derive exact results for the KPZ interface in one dimension. Interestingly enough, the 
analysis of some of its properties makes contact with random matrix theory, asymmetric exclusion 
processes, and other celebrated problems in statistical physics and mathematics. As soon as the dimension of the 
substrate goes beyond one, the exponents are no longer easy to derive, and no renormalisation 
group strategy has been fully successful yet.

All in all, the KPZ equation presented 
new analytic and experimental challenges which fostered the derivation of a long list of very interesting results. On the 
mathematical side, Martin Hairer was attributed the Field Medal  in 2014 for his formal studies of this equation~\cite{Hairer}. 
Kazumasa Takeuchi obtained the IUPAP Young Scientist Award 2013 for his experiments on growing interfaces in liquid crystal turbulence.
Thanks to the sufficient statistics gathered with this system, K. Takeuchi  managed  to study 
height distributions and correlation functions and with them he demonstrated 
the KPZ universality class~\cite{Takeuchi}. 

\subsection{Stochastic resonance}

The Stochastic Resonance phenomenon~\cite{Benzi} is exemplifed by the rather simple non-linear Langevin equation:
\begin{equation}
{\rm d}_t T(t) =
\underbrace{T(t)(a-T^2(t))}_{\rm \textcolor{black}{non-linear}}+\underbrace{A \cos (\Omega t)}_{\rm \textcolor{black}{periodic}}  + 
\!\!\!\! \underbrace{\xi(t)}_{\rm white \ noise}
\end{equation}
with $a>0$.
In the absence of the periodic perturbation, $T(t)$ relaxes to one of the minima of the double-well potential at the origin of the 
deterministic force in the right-hand-side and makes random jumps over the barrier via thermal activation to the other one,
back, and so on and so forth. With the addition of the periodic perturbation, and for not too restrictive choices of the 
strength of the noise, a cooperative effect between the non-linear relaxation and the external forcing emerges. It results 
in a strong response in the power spectrum at the frequency $\Omega$, the resonance, 
which corresponds to a noisy  nearly periodic motion 
with amplitude $2\sqrt{a}$. In the words of the authors ``It is conceivable that this new type of resonance might play a role in explaining the 
$10^5 $ year peak in the power spectra of paleoclimatic records'' which corresponds roughly to the alternation between 
glacial and interglacial stages.

\subsection{Super-symmetry and dimensional reduction}

In a couple of other fundamental papers, Parisi and Nicolas Sourlas~\cite{ParisiSourlasPRL,ParisiSourlas} 
introduced a super-symmetric representation of stochastic equations which proved itself not 
only elegant but also very useful. 

On the one hand, this mapping gave a very nice justification of the  
(perturbative) dimensional reduction of the Random Field Ising Model (RFIM)~\cite{ParisiSourlasPRL}. 
The latter states that the most infrared-divergent diagrams in the $d$ dimensional RFIM are equal
to the same diagrams in the model without magnetic fields in $d-2$ dimensions. 
The Parisi-Sourlas proof goes as follows. The  field $\phi$ correlation functions 
at the tree (most divergent) level are written in terms of random field $h$ averages over 
the solutions of the classical stochastic equation $-\nabla^2 \phi + V'(\phi) + h = 0$.
The correlation is then re-written as a functional integral over the field $\phi$, with the 
classical equation being imposed with a delta function times a determinant. The latter 
is expressed as an integral over auxiliary fermionic fields and the remarkable 
fact is that the ensuing field theory has a super-symmetry.  A more compact 
expression is obtained introducing the super-space with two added anti-commuting 
coordinates and once this done the idea behind
dimensional reduction is that the super-space in which the super-symmetric theory 
lives is equivalent to $d-2$ real space.

In their second paper~\cite{ParisiSourlas}, Parisi and Sourlas extended the connection to other 
super-symmetric field theories and classical stochastic equations. 
They also studied the spontaneous breaking of super-symmetry and showed that 
it is related to the number of solutions of the stochastic equations.

\section{Computer design \& observational methods}
\label{sec:computers}

Parisi actively participated in the construction of special purpose computers. 

\subsection{Lattice QCD} 

In the 80s he was much involved in the 
early steps and further development of the ``Array processor with emulator'' (APE) computer built to perform Monte Carlo simulations for lattice QCD. 
This was a collaboration based at the INFN Sections of Roma ``La Sapienza'', Roma “Tor Vergata”, Pisa, Bologna, and Padova,
which was leaded by Nicola Cabibbo, who gave a strong scientific
guidance to the project, and several remarkable junior scientists, who are at present 
members of the physics departments of Roma I and Roma II, like Enzo Marinari and Gaetano Salina, respectively. 
APE was a parallel SIMD machine, with architecture optimised for complex number arithmetics, and its own Fortran-like language. 
Parisi masterminded the machine architecture, and was the principal author of the compiler, the random number generator, highly optimised lattice QCD codes, {\it etc}. Three generations of APE computers testify his influence, namely the 1 Gflop APE (1985-1987), 
the 100 Gflop APE100 (1989-1994) and the 1 Tflop APE mille (1995-2000). 
In 1991 APE100 was the most powerful supercomputer in the World. DESY and Orsay acquired versions of this machine, putting European 
computational power in Lattice Field Theory on a par with that of the US and Japanese groups.

Parisi used the APE machines in a multitude of studies of fundamental non-perturbative aspects of QCD. Some of the 
issued analysed by APE are
glueball masses and the string tension~\cite{Albanese}, the Hadronic mass spectrum, the QCD deconfining phase transition, 
mesonic and meson-nucleon scattering lengths.
Some rather bold approximations had to be adopted and one of them was the {\it quenching} of fermions (this name given by Parisi 
was motivated by his knowledge of disordered systems)~\cite{Fucito}, innovative techniques were developed so 
as to reliably extract physical information (e.g. “APE smearing”). These early achievements paved the way to 
the present-day attainment of lattice QCD high-precision results, which are nowadays habitually used by experimentalists and phenomenologists 
using the Standard Model and beyond. 

\subsection{Finite dimensional spin-glasses}

More recently, he was the leader of a Roma - Ferrara - Badajoz - Madrid - Zaragoza collaboration which built and extensively used 
a series of computers (SUE and various Janus generations) to perform Monte Carlo simulations of Ising spin - glasses mostly in finite 
dimensions~\cite{Baity-Jesi}. The purpose of these studies was to put the replica symmetry breaking picture to the test, and 
study the relaxation after quenches into the ordered phase like in the experimental protocol used in different laboratories.

\subsection{Roman starlings}

The Piazza del Cinquecento facing the Roma Termini train station exhibits a notable show from November and until February, circa: 
the  dance of  starlings' flocks
at around sunset before they roost in the numerous nearby trees. 
Parisi, with the invaluable contribution of Andrea Cavagna and Irene Giardina, set in place over the roof of Palazzo Massimo, Museo Nazionale Romano,  
an observational system  which, using stereometric and computer vision techniques, allowed them to reconstruct the three dimensional 
trajectories of individual birds. The gathered unprecedented data-sets for around 3000 individuals were the starting point for a series of analysis which allowed 
to  clarify several issues about the organisation of the flocks. For instance, one the first conclusions reached by the team was that 
birds follow their neighbours using a topological (instead of metric) distance basically coordinating on average with a fixed numberof 
them~\cite{Ballerini}. The number extracted from the data analysis,  six to seven, is significantly smaller than the number of visually unobstructed 
neighbours around each bird. Research on animal collective motion continued in Rome 
under the supervision of Cavagna and Giardina until present.

\section{Conclusions}

Parisi trained a huge number of students, post-docs and young researchers, building an international school of research with 
most important impact in Italy and France. He used a rather uncommon way of doing research at the time, 
applying all available tools to the problems he wanted to solve, be them analytical, numerical or phenomenological,  all combined with
his great  intuition.
All of us, who worked with him at some point, were exposed to this style of reseaerch, and tried to imitate it with more 
or less success. 

\begin{figure}[h!]
\begin{center}
\hspace{0.7cm}
\includegraphics[scale=0.42]{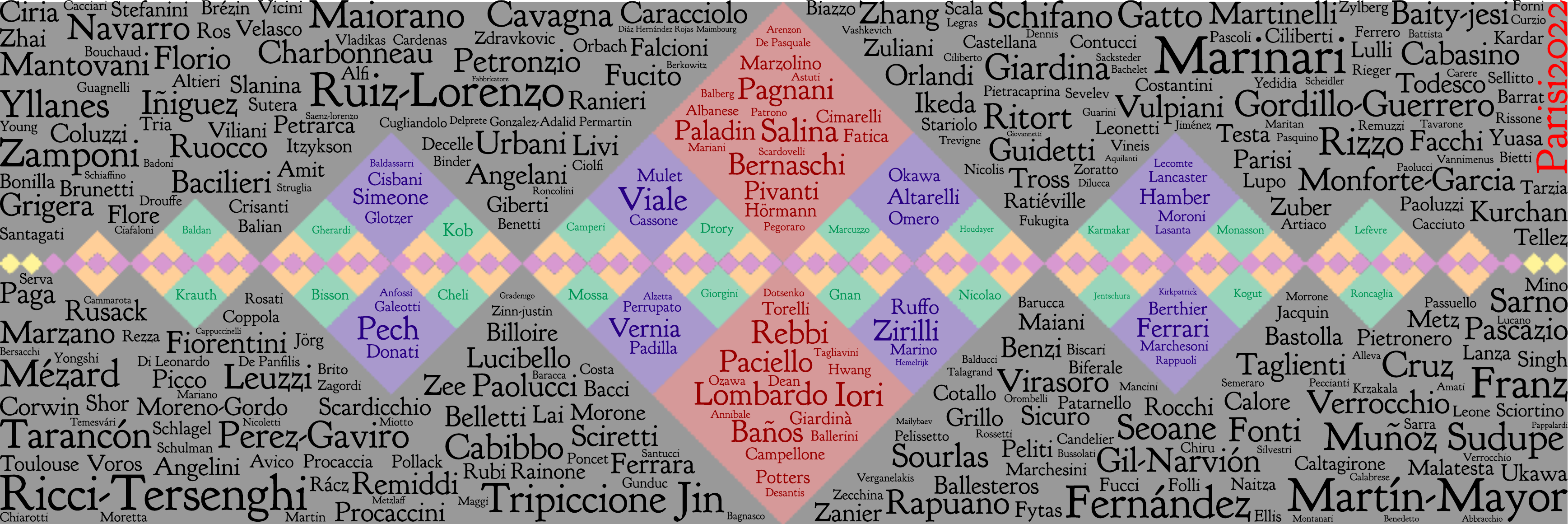}
\end{center}
\caption{Giorgio Parisi's collaborators as of May 2022.}
\label{fig:co-authors}
\end{figure}

G. Parisi has published papers with around 380 co-authors 
and counting. For his 70th birthday, his former associates Maria Chiara Angelini, Gabriele Sicuro and 
Pierfrancesco Urbani prepared a first version of the co-author 
map   and keep it updated. The final version, as of May 2022, is shown in Fig.~\ref{fig:co-authors}.

\vspace{0.5cm}

In 2018, the  40th anniversary of Parisi's first paper on Replica Symmetry Breaking (RSB) was celebrated in Rome
with a conference in which many of the main actors in the use and development of these ideas presented
their work. A book with a collection of chapters in which applications of RSB to different
fields will soon be published~\cite{RSB40}. Though completely useless from a practical point
of view the spin-offs of the theoretical work done to describe spin-glasses has been utterly important
in very different areas.

\vspace{0.5cm}

\noindent
{\bf Acknowledgements}
I warmly thank F. Bouchet, I. Giardina, G. Parisi, J. J. Ruiz-Lorenzo, G. Salina, N. Sourlas, A. Vladikas, F. Zamponi and J-B Zuber for helping me 
gather necessary information that is used to construct several passages of this text.

\newpage

\bibliographystyle{unsrt}
\bibliography{references}

\end{document}